\newcommand{\be}[0]{\begin{equation}}
\newcommand{\ee}[0]{\end{equation}}
\newcommand{\bea}[0]{\begin{eqnarray}}
\newcommand{\eea}[0]{\end{eqnarray}}

\documentclass[12pt]{article}

\usepackage{graphicx}
\usepackage{psfrag}

\usepackage{txfonts}                   

\begin{document}
\large
\hfill\vbox{\hbox{DCPT/05/122}
            \hbox{IPPP/05/61}}
\nopagebreak

\vspace{2.0cm}

\begin{center}
\LARGE
{\bf Scalars in the hadron world:\\ the Higgs sector of the strong interaction}

\vspace{0.8cm}

\large{M.R. Pennington} 

\vspace{0.5cm}

{\it Institute for Particle Physics Phenomenology,\\ Durham University, 
Durham, DH1 3LE, U.K.}\\
     
\vspace{3.0cm}

\end{center}

\centerline{\bf Abstract}

\vspace{0.3cm}

\small

Scalar mesons are a key expression of the strong physics regime
of QCD and the role   condensates, particularly 
$\langle q{\overline q} \rangle$, play in breaking chiral symmetry.
 What new insights have been provided by recent experiments on $D, D_s$ and $J/\psi$ decays to light hadrons is discussed. We need to establish whether all the claimed scalars $\sigma$, $\kappa$, $f_0(1370)$, etc., really exist and with what parameters before we can meaningfully speculate further about which is 
transiently ${\overline q}q$, ${\overline{qq}} qq$, multi-meson molecule or largely glue.

\vspace{1.cm}

\normalsize

\vspace{5.5cm}

\noindent Invited talk at the {\it International Conference on QCD and Hadronic Physics}, Beijing, June 2005.

\newpage

\parskip=2mm
\baselineskip=5.2mm

\section{Structure of the QCD vacuum}

\noindent This meeting celebrates QCD as the theory of the strong 
interactions. Many of its successes over the past 30 years stem from its 
remarkable property of asymptotic freedom. This means that at short distances 
we can use perturbation theory to describe hard scattering processes, make 
predictions and find they agree with experiment. But the bulk of hadronic 
phenomena
involve interactions typically over the size of a proton and over such distances 
the interactions are strong. For these, QCD is far more difficult to solve 
and so we have to use a mixture of modelling, approximations and guidance 
from experiment.
The strong coupling regime of QCD is responsible for the spectrum of hadrons, 
which in turn is intimately related to confinement and for light hadrons to 
the property of chiral symmetry breaking.  These are the topics this talk encompasses. 

QCD is rooted in the spectrum of hadrons and the relation of this to the quark 
model. Ideas we will question here. Nevertheless, the bulk of hadrons do fit
 into quark model multiplets. For ${\overline q}q$ states, the template is that for vector mesons
 formed 
by combining a quark and an antiquark in a spin-one ($S=1$) system with no 
orbital angular momentum ($L=0$) between them. Then with three flavours of 
quark, we expect 9 vector mesons. Nature tells us that the states in the 
middle of the multiplet with
zero third component of isospin, the $\rho$, $\omega$ and $\phi$ are states of 
definite quark flavour. This we infer from their pattern of decays and their 
masses. Almost all mesons fit into such ideally mixed multiplets. However, 
it is well known that there are far more scalars than can fit into one such 
multiplet with $S=L=1$. We have the $f_0(400-1200)$, $f_0(980)$, $f_0(1370)$, 
$f_0(1500)$, $f_0(1710)$, $K_0^*(1430)$, $a_0(980)$, $a_0(1430)$ and perhaps 
even others~\cite{PDG}.
Why does it matter that these don't fit our quark model template? Why indeed  
is there a special  talk at this conference on the scalars? This is because 
the scalars are special. They are the Higgs sector of the strong interaction 
and directly reflect the structure of the QCD vacuum. 

In QED, the vacuum, through which 
an electron in an atom moves, is filled with electron-positron pairs. The effect
 of these can be calculated in perturbation theory and so determine the 
fine structure of  atomic spectra. Because of asymptotic freedom, the QCD 
vacuum appears similar over short distances. Quarks move through a sea of 
${\overline q}q$ pairs and a shoal of gluons, but, so strong are the 
interactions between quarks and gluons over long distances, these change the 
nature of the vacuum. The long range correlations form condensates of quarks, 
antiquarks and gluons.

The symmetries of the hadron world are a reflection of QCD. 
While isospin symmetry follows from the near equality of {\it up}
 and {\it down} quark masses, it is their current masses that enter the 
QCD Lagrangian. These are only a few MeV and so
much less than the natural scale of QCD, {\it viz.} $\Lambda_{QCD}$, which is 
100-200 MeV. Thus to a good approximation the {\it up} and {\it down} 
quarks can be regarded 
as massless.  The QCD Lagrangian then has a bigger symmetry. 
Left-handed {\it ups} 
and {\it downs} can be interchanged independently of those spinning 
right-handedly 
and vice-versa. Consequently, the QCD Lagrangian has an $SU(2)\otimes SU(2)$ 
symmetry, and in as much as {\it strange} quarks are light, 
it has 
a more approximate $SU(3) \otimes SU(3)$ symmetry. These symmetries would be mirrored
 in hadron interactions too, but we do not see them. 
Scalars and pseudoscalars, vectors 
and axial-vectors do not have simply related interactions and masses. This is 
because the symmetry is broken. Nambu  proposed a model~\cite{nambu} of this even before 
QCD was discovered. If the symmetric potential generated by scalar and 
pseudoscalar interactions was not like a bowl, but Mexican hat shaped, then 
nature would spontaneously break the symmetry. Let us call the  scalar  
$\sigma$ and identify the pseudoscalar with the $\pi$. The scalar has a non-zero vacuum 
expectation value. The physical particles that correspond to the quantum 
fluctuations about the minimum of the potential (the vacuum) feel no resistance in the $\pi$ direction and 
the pions are massless, while the oscillations in the $\sigma$ direction go up and down the parabolic sides of the hat and so have mass~\cite{nambu,MRP}. This spontaneous breaking of chiral symmetry is just like the magnetisation of a ferromagnet. The pions, the lightest of all hadrons, are the Goldstone bosons, while the scalars are the Higgs sector. Their mass gives mass to all light hadrons. 
At the QCD level, the chiral symmetry is dynamically broken by the formation of condensates.

 In the last few years we have learnt that this breaking is dominated by the
non-zero value of the ${\overline q}q$ condensate with a scale
of $\sim -(250 {\rm MeV})^3$. We learn this phenomenologically using QCD sum-rules~\cite{MRP}, theoretically from calculations of quark mass functions in the chiral limit which are made possible by using the Schwinger-Dyson equations in the continuum~\cite{MRP}, and most recently experimentally by accurate determinations of low energy $\pi\pi$ scattering from $K_{e4}$ decay~\cite{pislak} and  studies of pionium~\cite{dirac}.
These confirm that the expansion of the pion mass in terms of quark mass is dominated by the very first term~\cite{colangelo1}, as in the Gell-Mann-Oakes-Renner relation~\cite{GMOR}:
\begin{equation}
m_{\pi}^2\,f_{\pi}^2\;=\;-\,(m_u\,+\,m_d)\;\langle\, O\,\mid \,{\overline q} q\,\mid\,O\, \rangle\;+\; O(m_q^2)\; ,
\end{equation}
where $f_{\pi}$ is the pion decay constant. This dominance by the first term tells us that Nambu's ferromagnetic analogy is very close to what happens in the real world. Consequently, it is natural to ask what is the scalar field, whose non-zero vacuum expectation value breaks the chiral symmetry and whose mass reflects the constituent mass of light quarks?
What is the chiral partner of the pion, and to a more approximate extent what is the chiral partner of the kaon? What are the $\sigma$ and the $\kappa$? Are they just one field or a collection of particles, $f_0$'s and $K_0^*$'s, seen in experiment?
\section{Scalars in scattering experiments}
Let us first briefly review what we know about the scalars from the classic meson-meson scattering experiments  and then we will turn to what the more recent
decay results tell us. The easiest states to identify unambiguously are the strange ones. These we learn about from high energy $K\pi$ production in $K p$ collisions, which at small momentum transfers are dominated by one pion exchange. From the famous LASS experiment~\cite{lass} of twenty years ago, we see that $K\pi$ scattering is dominated by the spin-1 $K^*(892)$ and the spin-2 $K_2^*(1430)$ each seen in the appropriate partial wave with the rapid phase variation expected of a resonating wave. Under these the spin-0 wave has a broad bump rising to the unitarity limit with the slower phase variation expected of a state of 250 MeV width. This is the $K_0^*(1430)$. These are the $I=1/2$ scalar mesons.

\begin{figure}[t]
\begin{center}
\includegraphics[width=10.cm]{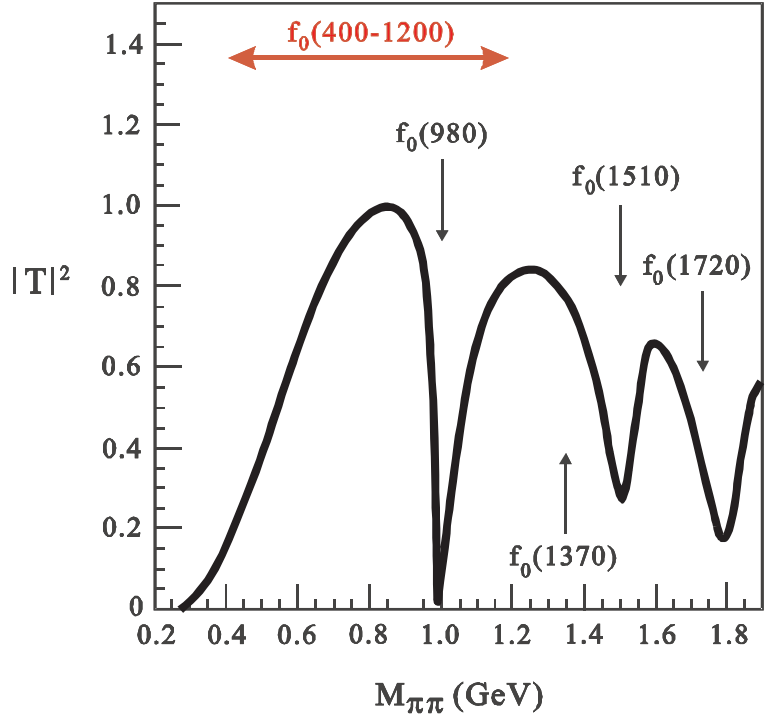}
\end{center}
\vspace{-4mm}
\caption{\leftskip 1.5cm\rightskip 1.5cm{A sketch of the square of the modulus of the $I=0$ $\pi\pi$ $S$-wave 
amplitude, from Zou~$^{10}$.}}
\end{figure} 
The $I=1$ states are found by partial wave analysing channels like
$\pi^- p \to (\pi^0 \eta)n$ studied by GAMS~\cite{GAMS}. Again this process is dominated  at small momentum transfers by pion exchange and reveals two states the $a_0(980)$ and $a_0(1430)$. The fact that there are two low mass isotriplets, may suggest that there should be another pair of isodoublets too. The LASS results on $K^-\pi^+$ scattering start at 825 MeV, 200 MeV above threshold, and there has been intense speculation that there may be another scalar state, the $\kappa$, closer to threshold. We will discuss this possibility shortly.

The $I=0$ channel is studied in $\pi\pi$ scattering. The $J=0$
cross-section shown in Fig.~1 reveals a series of peaks and dips~\cite{zou}, none of which is
describable by any simple Breit-Wigner form. The first dip is strongly
correlated with the onset of the $\pi\pi\to{\overline K}K$ channel and
marks the $f_0(980)$. From Fig.~1 we see many overlapping isoscalar states. Any channel claiming to see any of these must include a
description of all of them (though they may well appear with different strengths in different
processes). 
They are inextricably mixed with each other and with the thresholds to
which they couple. One cannot arbitrarily pick and choose and
describe one or two with simple Breit-Wigner forms and ignore the
others. 

\begin{figure}[t]
\begin{center}
\includegraphics[width=14.4cm]{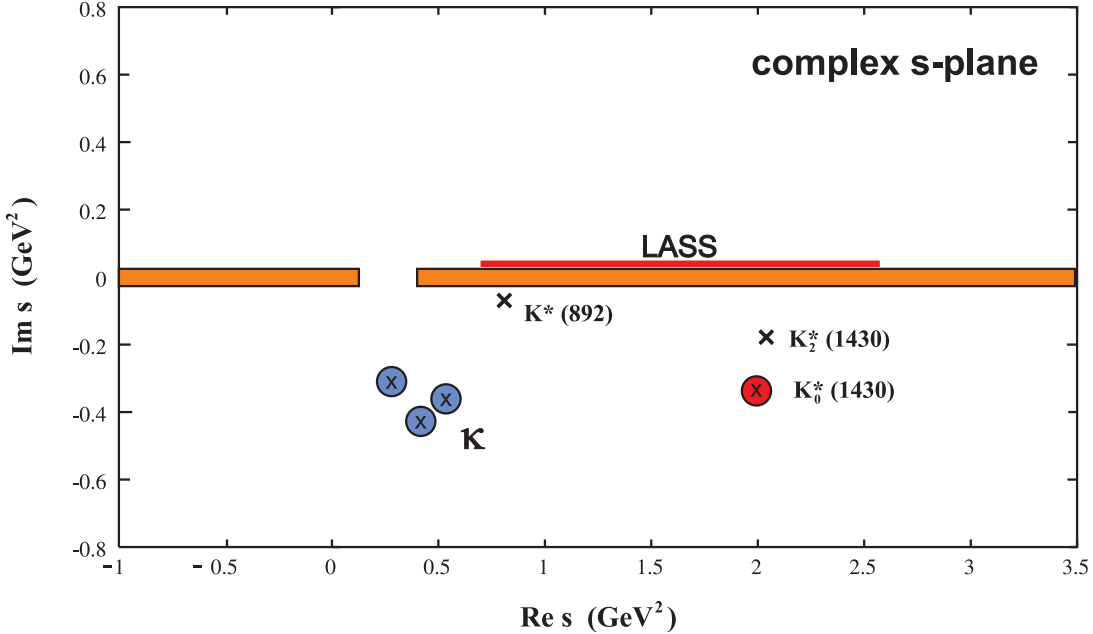}
\end{center} 
\vspace{-2.mm}
\caption{\leftskip 1cm\rightskip 1cm{The complex $s$-plane for $K\pi$ scattering, showing the cut structure,
and the nearby poles on the unphysical sheet. The possible positions of the $\kappa$-pole are from the listing in Ref.~1. } }
\end{figure} 

Fig.~1 indicates 5 low mass states. These could populate two
nonets and still leave one that might be a glueball. 
Before speculating about the nature of such states, let us
first illustrate why we need precision results to be certain that they
all exist as poles in the complex energy plane. This need is readily
understood if we consider the complex $s$-plane for $K\pi$
scattering. The amplitude (whether the forward full amplitude, or
individual partial wave amplitudes) has a right hand cut produced by
direct channel dynamics, and a left hand cut generated by crossed
channel exchanges. In Fig.~2 is shown this plane. Along the top of the
right hand cut, the region accessible in the LASS experiment is
delineated. 
The determination of poles requires an analytic continuation, which is
only 
accurately possible near to this region. Thus we can find the
$K^*(892)$, $K^*_2(1430)$ and even the wider
$K^*_0(1430)$, but whether there is a low mass $\kappa$ or not depends
on continuing experimental results into \lq\lq unknown'' territory. There is certainly no such pole
with Re$s\,>\,0.7$ GeV$^2$ as shown in Ref.~11.
However,
different parametrisations of the data do yield a pole~\cite{PDG} deep in the
complex plane below Re$s\,< 0.6$ GeV$^2$. But this continuation enters regions
just as close to the left hand cut as to where we have scattering
data. Consequently, parametrisations that do not have the correct left hand
cut analyticity, like simple $s$-channel Breit-Wigner forms, are highly
suspect. The use of chiral perturbation theory to constrain the
continuation is clearly an improvement, but the uncertainties in how to sum all orders are so large that we cannot be
sure that the pole is not really much deeper.   It is then an open question
whether such a pole is meaningfully present. 
\begin{figure}[t]
\begin{center}
\includegraphics[width=14.2cm]{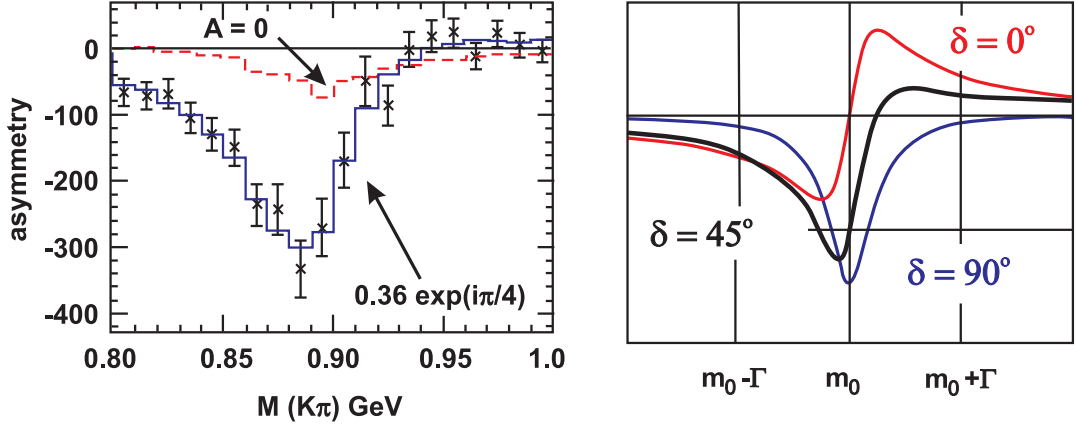}
\end{center}
\vspace{-3.5mm} 
\caption{\leftskip 1cm\rightskip 1cm{Asymmetry distribution in $K\pi$ invariant mass from the FOCUS experiment~$^{12}$. In the left hand figure the dashed line represents a simulation with no interfering $S$-wave. The solid line is with an $S$-wave of constant magnitude and phase of $\pi/4$. The plot on the right shows how the asymmetry (defined in Ref.~12)
changes for three different choices of $S$-wave phase.}}
\vspace{-1mm}
\end{figure} 

\section{Scalars in decays and in the complex energy plane}
Fresh light is shed on the nature of the scalars by
heavy flavour decay results --- on $J/\psi$ and $D$-decays in particular. Let us start with the semileptonic decay $D^+\to (K^-\pi^+)\, \mu^+\nu_{\mu}$, which FOCUS~\cite{focusdl4} has measured in the 700-900 MeV $K\pi$ mass range. In this region the strong interaction component is dominated by the $K^*(892)$. If this was all there was then there would be no forward-backward asymmetry in the $K\pi$ rest frame. However, there is a strong asymmetry, shown in Fig.~3. This tells us that in addition to the $P$-wave there is an $S$-wave $K\pi$ interaction with phase of about $45^o$. This is just what we would expect from a dominant $I=1/2$ final state interaction with phase given by the LASS results~\cite{lass} as required by Watson's theorem. The phase is not $90^o$ (Fig.~3), which  a Breit-Wigner for a low mass $\kappa$ would suggest. So far $D_{l4}$ decay has only been studied in the 700-900 MeV region: the undetected neutrino causes the difficulty. Nevertheless, higher statistics from $B$-factories have the potential  to allow the  extraction of the $S$-wave $K\pi$ phase in different charge channels down towards threshold. This would provide a key piece of the jigsaw puzzle of low energy hadron dynamics.

We now turn to the potentially more complicated multi-hadron $D$-decays such as $D^+\to K^-\pi^+\pi^+$. With well established resonances dominating the $K^-\pi^+$ interaction with spin-1 and spin-2, one can use knowledge of these to determine the $S$-wave amplitude, both magnitude and phase in $K\pi$ mass bins across the Dalitz plot. These have been determined from the E791 data by Brian Meadows~\cite{meadows}. These show a variation of phase that is not wholly attributable to the $I=1/2$ $S$-wave. As illustrated by Laura Edera and I~\cite{edera} one can use the way the magnitude and phase change through the region of elastic unitarity to separate the $I=1/2$ and $I=3/2$ $S$-waves. An important feature of these results is that they already provide phase information down to $K\pi$ threshold, even without $D_{\ell 4}$ results. While the E791 data are not yet of sufficient precision to determine the phase within a fraction of a degree, higher statistics results from BaBar hold out the prospect of being able to constrain the $S$-wave amplitude down towards threshold at least as well as the analytic continuation of the LASS results~\cite{lass} allows and hence help to determine whether a $\kappa$ resonance really exists as a pole in the complex energy plane or not.

In the $I=0$ sector it has long been known that the shape of the $\pi\pi$ mass distribution reflects the production mechanism. In Fig.~1, we have already seen the shape of the cross-section in $\pi\pi\to\pi\pi$ scattering with its broad low mass enhancement and deep dip at 1 GeV. For central $\pi\pi$ production in pp collisions at the ISR or $\pi^-p$ interactions at WA102, the $\pi\pi$ mass distribution is more peaked just above threshold  because in this process there is no Adler zero. The reaction is essentially Pomeron-Pomeron scattering to $\pi\pi$ which is controlled by $\pi$ exchange as discussed long ago in Ref.~15.
There the $f_0(980)$ appears as a shoulder. In contrast, in
$J/\psi \to \phi(\pi\pi)$ and $D_s\to\pi(\pi\pi)$, where hidden strangeness is the source of the $\pi\pi$ system, the $f_0(980)$ appears as a striking peak.
This preference for coupling to ${\overline s}s$, or equivalently virtual ${\overline K}K$, would in a conventional quark model multiplet indicate that the $f_0(980)$ is predominantly the ${\overline s}s$ state. However, as is well known, it is then difficult to understand how it can be degenerate in mass with the $a_0(980)$, which would have no strange quarks. This, of course, begs a very basic question. When are  hadrons and their underlying quark model states closely identified?

 We know the $\phi$ meson is not just ${\overline s}s$, but contains within its Fock space contributions from ${\overline K}K$ and $\rho\pi$. It is through these components that it decays. However, these multi-meson components are small. They affect the parameters of the vector meson little. Consequently, we identify the $\phi$ with ${\overline s}s$ and the $\rho^+$, for instance, with
$u{\overline d}$, even though they contain ${\overline K}K$ and $\pi\pi$ components, respectively. Vector states coupling to two pseudoscalars require $P$-wave interactions that are naturally suppressed near threshold. The fact that additional quark loops contribute little is, of course, totally in keeping with the $1/N_c$ expansion, which would make these loops higher order. Unquenching is unimportant. In contrast, for the scalars, like the $f_0$ or $a_0^+$, even if they are seeded by ${\overline s}s$ and $u{\overline d}$, respectively, quark loops do matter. They drastically affect the states
and give the $f_0$ and $a_0$ each a ${\overline K}K$ component~\cite{weinstein,vanbeveren} that is as big as $\sim 40\%$. Thus whatever their intrinsic make-up they have a significant long range component like a ${\overline K}K$ molecule. Unquenching is important in the scalar sector. There is no $1/N_c$ suppression. Indeed, this means the OZI rule does not hold here. Not only do ${\overline s}s$ and ${\overline u}u$, ${\overline d}d$ communicate, but the presence of these resonances near 1 GeV even enhances this communication~\cite{moussallam}. This must be a clue to the flavour structure of the QCD vacuum.

Counting states determines whether the inevitable 
${\overline K}K$ nature of the $f_0,\,a_0(980)$ is the result of an
 intrinsically
molecular structure as claimed by Weinstein and Isgur~\cite{weinstein},
or whether they are just two examples of primarily 4 quark configurations.
In the latter case there would be two low mass scalar multiplets, one of ${\overline q}q$ and the other ${\overline {qq}}qq$, as proposed by Jaffe~\cite{jaffe}, by Schechter and 
collaborators~\cite{schechter} and more recently by Maiani {\it et al.}~\cite{maiani}, amongst others. The idea of systems built of scalar diquarks of different flavours, like $[ud]$, $[us]$ and $[ds]$, has recently received renewed interest with the many discussions of the $\Theta^+(1540)$ the putative pentaquark baryon.
This has highlighted the possibility of a nonet of tetraquark states
made of these three light types of diquarks. While in a conventional
${\overline q}q$ nonet, the ${\overline s}s$ state is the heaviest, in
a tetraquark multiplet there is a degenerate isotriplet as well as an
isosinglet built of $[{\overline {ns}}] [ns]$ with $n=u, d$. This would
  naturally explain the degeneracy of the $a_0(980)$ and $f_0(980)$. In
  such a modelling, the ${\overline q}q$ nonet is heavier and would
  incorporate the $K^*_0(1430)$, $a_0(1430)$, and two of the
  $f_0(1370)$, $f_0(1500)$ and $f_0(1710)$ leaving one to be a
  glueball. Different schemes favour one or the other to be largely gluish as we have heard at this meeting~\cite{FEC}. 
The lighter tetraquark multiplet then needs strange isodoublets, the $\kappa$, and lightest of all a $[{\overline {ud}}][ud]$ ($f_0$ or) $\sigma$. However seductive this picture, further discussion is not fruitful until we have established the existence of the complete set of states: in particular $\sigma$, $\kappa$ and $f_0(1370)$ as unequivocal states in the spectrum. Within a $1/N_c$ framework, as $N_c\to \infty$ the ${\overline q}q$ states become stable, while the tetraquarks merge with the meson-meson continuum~\cite{jaffe2,pelaez}. As we have already noted the real world of scalars with $N_c=3$ is distinct from the $N_c\to \infty$ limit. Thus, it is 
real world meson-meson scattering, as seen in Fig.~1, with  whatever states it encompasses, that determines 
the ${\overline q}q$ condensate that drives chiral symmetry breaking, and {\it vice versa}.

Indeed in a quite different scenario, Minkowski and Ochs~\cite{minkochs} have proposed that it is the lightest scalar, the $f_0(400-1200)$, that is the glueball (or ``red dragon'' as they call it). $J/\psi$ decay has often been regarded as a source of glue.
Long ago both the Mark III and DM2 experiments found $J/\psi\to\omega (\pi\pi)$ decay has  a low mass $\pi\pi$ enhancement. 
Statistics were not sufficient to allow a partial wave separation. 
With the higher event rates now available at BEPC, such  a separation 
becomes possible. This confirms that the effect is indeed due to $S$-wave $\pi\pi$ 
interactions. BESII have fitted this with  Breit-Wigner forms and found~\cite{BESii} 
$M=(541\,\pm\,39)$ MeV and $\Gamma=(504\,\pm\,84)$ MeV.
Before this, E791 analysed $D^{\pm}\to\pi^{\pm}\pi^+\pi^-$ decay and
 found a  low mass enhancement too. Their  Breit-Wigner fit~\cite{e791} yields a similar $M=(478\,\pm\,29)$ MeV but a narrower $\Gamma=(324\,\pm\,23)$ MeV. The
 FOCUS experiment has comparable $D$-decay data and  a similar fit reveals
 a $\sigma$ with similar parameters. As emphasised earlier the scalars seen in Fig.~1 are
inextricably linked, yet the E791 fit~\cite{e791b} to $D_s\to 3\pi$ introduces a Breit-Wigner description for a state
at 1434 MeV with width of 173 MeV, which is neither the $f_0(1370)$ or $f_0(1500)$, seen in Fig.~1. Similarly, Belle 
fit ${\overline D}^0\to K_s\pi^+\pi^-$ with not only a low mass $\sigma$, but another near 1 GeV in addition to the $f_0(980)$~\cite{Belle} --- again not known in $\pi\pi$ scattering.
However, FOCUS~\cite{malvezzi} have also analysed
 their data on $D\to 3\pi$ in a way consistent with the $\pi\pi$ scattering 
results
 shown in Fig.~1 and found an even better fit. This does not mean that
 a low mass $\sigma$ is not needed, only that the $\pi\pi$ final state
 interactions are consistent with elastic scattering. Such a broad bump is not necessarily resonant, 
even if it can be fitted by a Breit-Wigner form! One has to
 show that there is indeed the corresponding phase variation. For the
 FOCUS data tests of this are still going on. However, for the BESII
 results on $J/\psi$ decay,  Bugg has analysed these in Ref.~31.
Using the interference with the crossed bands of $\omega\pi$ 
interactions in both 
$S$ and $D$-waves that produce the $b_1$, he has shown that the $\pi\pi$ 
$S$-wave has a phase variation totally consistent with results on $\pi\pi$ 
scattering, Fig.~4. As the parametrisation of this by Anisovich and Sarantsev~\cite{AS}, for instance, illustrates perfectly, this does
not require a low mass pole in the complex plane.
\begin{figure}[h]
\vspace{2.mm}
\begin{center}
\includegraphics[width=10.cm]{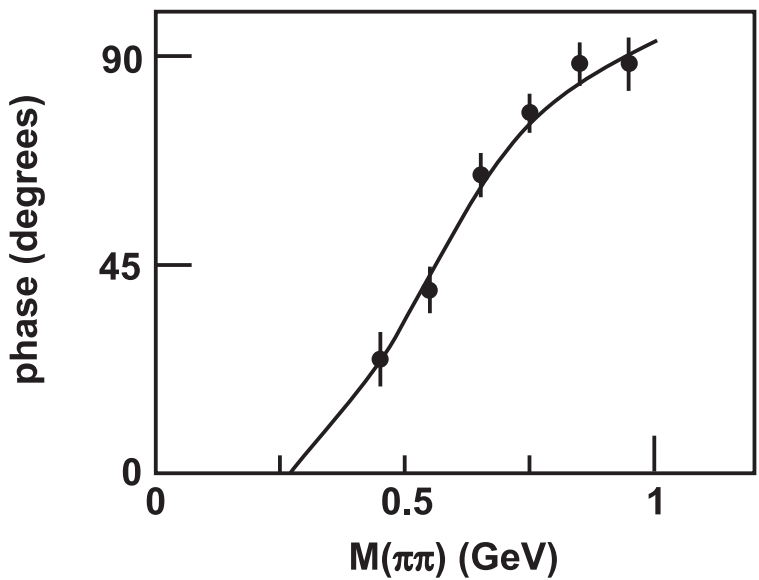}
\end{center} 
\vspace{-5mm}
\caption{\leftskip 1.cm\rightskip 1cm {Phase of the $\pi\pi$ $S$-wave in $J/\psi\to\omega(\pi\pi)$ determined from simultaneous fitting of the magnitudes and phases in each bin from Bugg~$^{31}$.}}
\end{figure} 

 Of course, simple Breit-Wigner forms for such a low mass enhancement presuppose a pole in the complex $s$-plane whether 
it is there or not. Moreover, they
do not incorporate the correct analytic
structure along the nearby left hand cut 
(just like Fig.~2 for $K\pi$ scattering), but for $\pi\pi$ scattering 
this is further constrained by three channel crossing symmetry.  
Fits by Zhou {\it et al.}~\cite{zhou} implementing crossing sum-rules find a pole at $M=(470\,\pm\,50)$ MeV with $\Gamma=(570\,\pm\,50)$ MeV.
 This compares very closely to the results of Colangelo, Gasser and Leutwyler~\cite{cgl}. 
Their implementation of the Roy equations determines near threshold $\pi\pi$ 
scattering as precisely as is presently possible. The $I=0$ $S$-wave is fitted with a 
parametrisation due to Schenk~\cite{schenk,cgl}. When continued into the complex plane 
this gives a pole within the region found by Zhou {\it et al.}~\cite{zhou}. 
However,
a pole so far from the real axis is not so readily determined. Unitarisations 
of chiral perturbation theory, like the inverse amplitude method, do find a 
low mass pole, as reviewed some time ago in Ref.~36,
but once again this does 
not prove that such methods provide the correct amplitude on the unphysical 
sheet. More work is needed to prove that this is not just the result of 
using too simplistic representations. We are however reaching a situation 
that precision data on heavy flavour decays when combined with correct 
analytic representations will determine whether the $\kappa$ and $\sigma$ (and $f_0(1370)$ for that matter) 
really exist as mesons in the spectrum of nature or not. Claims of states quoted above
which differ in 
width by hundreds of MeV depending on the representation used, 
do not yet prove that the true pole may  not yet be further away and even 
 off at infinity. The fields that are the Higgs of the strong interaction 
must remain an enigma for a little longer.
\vspace{3mm}

\noindent{\bf{Acknowledgments}}

\noindent It is a pleasure to thank Professors Kuang-Ta Chao and Chuan Li and their colleagues for the invitation to give this talk and to Bing-Song Zou, Ai-lin Zhang and Qiang Zhao for their very kind hospitality.
I  acknowledge partial support of the EU-RTN Programme, 
Contract No. HPRN-CT-2002-00311, \lq\lq EURIDICE''.

\end{document}